\documentclass[%
preprint,
 amsmath,amssymb,
pra,
]{revtex4-1}

\usepackage{graphicx}
\usepackage{dcolumn}
\usepackage{bm}
\usepackage{url}
\usepackage{amsmath}
\usepackage{amssymb}
\usepackage{array}
\usepackage{dcolumn}
\usepackage{bm}
\usepackage{mathrsfs}
\usepackage{subfigure}
\usepackage{color}

\usepackage{caption}
\usepackage{float}

\usepackage{hyperref}

\begin{document}


\title{Quantum Circuit Design for Training Perceptron Models}


\author{Yu Zheng$^2$, Sicong Lu$^{1}$, Re-Bing Wu$^{1\dag}$}
 \affiliation{$^1$Department of Automation, Tsinghua University, Beijing, 100084, China}


\affiliation{$^{2}$The Institute of Microelectronics, Tsinghua University, 100084, Beijing, China
}%

\date{\today}

\begin{abstract}
Perceptron model is a fundamental linear classifier in machine learning and also the building block of artificial neural networks. Recently, Wiebe {\it et al} (arXiv:1602.04799) proposed that the training of a perceptron can be quadratically speeded using Grover search with a quantum computer, which has potentially important big-data applications. In this paper, we design a quantum circuit for implementing this algorithm. The Grover oracle, the central part of the circuit, is realized by Quantum-Fourier-Transform based arithmetics that specifies whether an input weight vector can correctly classify all training data samples. We also analyze the required number of qubits and universal gates for the algorithm, as well as the success probability using uniform sampling, showing that it has higher possibility than spherical Gaussian distribution $N(0,1)$. The feasibility of the circuit is demonstrated by a testing example using the IBM-Q cloud quantum computer, where 16 qubits are used to classify four data samples.

\begin{description}
	\item[PACS numbers]
	42.50.Dv, 02.30.Yy
\end{description}
\end{abstract}

\maketitle

\section{Introduction}\label{sec:intruduction}
Recently, machine learning assisted by quantum computation, known as quantum machine learning (QML), has become an active field \cite{ma2017transforming,carleo2017solving,lu2017separability}. Many quantum algorithms have been proposed to learn from classical data samples, e.g., support vector machine\cite{cortes1995support}, principal component analysis \cite{jolliffe1986principal}, nearest-neighbor clustering \cite{cover1967nearest}, which all exihibit potentally important applications for their remarkable (ranging from polynomial to superpolynomial) algorithmic speedup and reduction of data storage.

As a simplest type of linear classifier, perceptron models played an essential role in the history of classical machine learning \cite{rosenblatt1958perceptron}. Mathematically, the perceptron model consists of a weighted sum of the input data and a nonlinear activation function, which mimics the biological reaction of a neuron to its environment. This feature endows perceptrons another important function, i.e., the building block of (feedforward and recurrent) artificial neural networks (ANN), the cornerstone of deep machine learning. Towards quantum perceptrons and quantum neural networks (QNN), various schemes have been proposed since 1990s. For examples, Lewenstein proposed that the unitary operator can be used to represent the weight vector \cite{lewenstein1994quantum}, but this scheme cannot be extended to large-scale quantum systems. Chrisley \cite{chrisley1995quantum} prensented a physial realization based on quantum optical setups, where the activeation of neurons is realized by double slits and the weights are trained by gradient algorithms as superposition states \cite{narayanan2000quantum}.

For quantum perceptrons, Ref.\cite{schuld2015simulating} proposed a circuit based on the phase estimation, which speeds up the calculation of the output but not the learning procedure. This is the reason why the technique of [11] is not used in our circuit. In \cite{ricks2004training}, Ricks presented an algorithm that quadratically speeds up the training process by Grover algorithm, and this route was recently developed by Wiebe \cite{wiebe2016quantum} based on Grover search in the so-called quantum version space perceptron model. 
Most of the above proposed schemes are still distant from practical implementations, to which a key step is to translate the abstract algorithms into quantum circuits. This is non-trivial as many arithmetic and logical operations are involved. In this paper, we will show how the quantum version space perceptron model \cite{wiebe2016quantum} can be decomposed and implemented by more elementary units of quantum circuits.
The remainder of this paper will be divided as follows. Section \ref{sec:model} will briefly introduce the version space model of quantum perceptron. Section \ref{sec:oracle} will decompose the circuit design into basic units and show how each of them can be implemented by circuits. In Section \ref{sec:experiment}, we will present the full quantum circuit for a example with four data vectors, and experimentally test it on the IBM-Q cloud quantum computer. Finally, we conclude and discuss the obtained results in Section \ref{sec:discussion}.

\section{quantum perceptron model}\label{sec:model}

Given a set of $K$ training data vectors $\mathcal{D}=\{({\vec x}_k,y_k),~k=1,\cdots,K\}$ with $y_k\in\{1,-1\}$ being the label of data vector ${\vec x}_k\in\mathbb{R}^n$, we wish to find a classifying hyperplane, whose normal vector is ${\vec w}\in\mathbb{R}^n$, that perfectly separates the two classes of data points labeled by $y=1$ and $y=-1$. The perceptron is associated with the following activation function,
$$y=f_{\vec w}({\vec x})=
\begin{cases}
1,& {\vec w}^T {\vec x}\geq 0\\
-1,& {\vec w}^T {\vec x}<0,
\end{cases}$$
and the goal of perceptron is to find a proper ${\vec w}$ such that $f_{\vec w}({\vec x}_k)=y_k$ for all $({\vec x}_k,y_k)$ in the training data set. Equivalently, the goal can be described as verifying whether 
\begin{equation}\label{}
f_{\vec w}({\vec x}_ky_k)=1,\quad~ \forall k=1,\cdots,K,
\end{equation}
which is easier to be implemented by a quantum circuit.
The classical training of a perceptron is usually done by iteratively updating the weights to reduce the number of misclassified data points until all data are correctly classified. In \cite{wiebe2016quantum}, the percptron model is transformed into the so-called version space, in which a data vector becomes a hyperplane and the weight vector becomes a vector. The feasible perceptrons are restricted to the hyper-pyramid intersected by the hyperplane, and hence the training of a perceptron can be done by randomly generating sufficiently many sampled vectors in the version space, and find the vectors that fall in the hyper-pyramid.

In Ref.\cite{wiebe2016quantum}, the success probability is proven to be approximately proportional to a separability measure $\epsilon$ (i.e., the distance between the two subsets that are to be separated) of the training data set, where the weight vectors are sampled according to a multi-dimensional Gaussian distribution. Note that the Gaussian distribution needs to be realized by a separate quantum circuit that can be very complicated. Here, we use the simplest Hadmard gates that produces a uniform distribution of sampling weight vectors. In the appendix, we show that the success probability has a similar scaling with that of Gaussian distribution when the weight vector is unifromly sampled from the unit sphere of the version space, and it can be higher when the dimension of the version space (or feature space) increases (see Fig.~\ref{fig:compare}). This qualifies the use of Hadmard gates as an alternative sampling circuit in our design.

\begin{figure}
	\center
	\includegraphics[width=0.8\columnwidth]{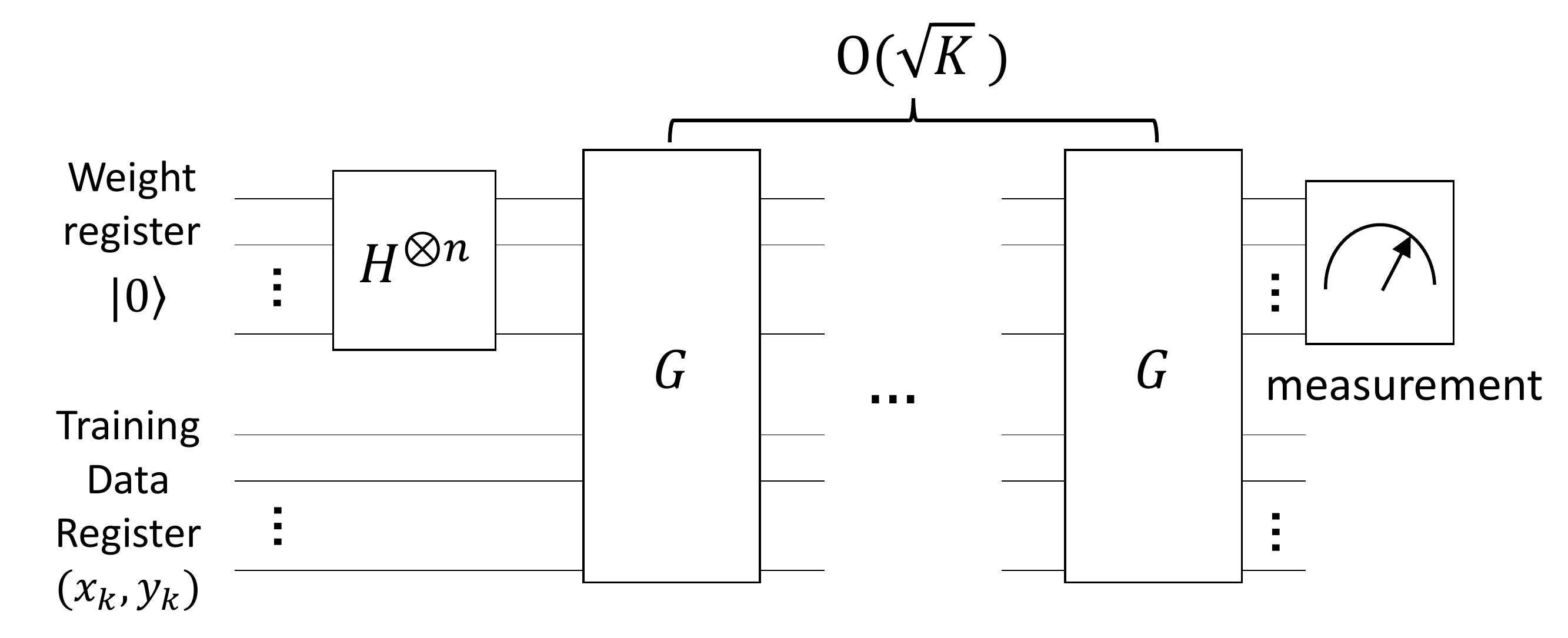}
	\caption{The quantum circuit for Grover search algorithm.}
	\label{fig:search}
\end{figure}

As illustrated in Figure~\ref{fig:search}, the training is done by a standard Grover search routine. The Hadmard gates generate a uniform superposition of $2^n$ samples of ${\vec w}$,
$|{\vec w}\rangle=2^{-n/2}\sum_{j=1}^{2^n}|{\vec w}_j\rangle,$
where $n$ is the number of qubits for encoding ${\vec w}$. The uniformly superposed samples of ${\vec w}$ are then fed into the Grover oracle, which flips the sign of the amplitude of the samples that can correctly separate all data points, i.e., $|{\vec w}_j\rangle\rightarrow -|{\vec w}_j\rangle$ if and only if the following Boolean function:
\begin{equation}
F({\vec w}_j)=f_{{\vec w}_j}({\vec x}_1y_1)\wedge  \dots \wedge f_{{\vec w}_j}({\vec x}_ny_n)=1.
\end{equation}

Afterwards, a phase reversing operation \cite{nielsen2002quantum} is carried out so as to amplify the amplitudes of those samples that correctly separate all data points. Suppose that there are $k$ feasible solutions, the Grover algorithm can find at least one solution in $\sqrt{K/k}$ iterations, where a classical search algorithm requires around $K/k$ iterations.

\begin{figure}
	\center
	\includegraphics[width=0.8\columnwidth]{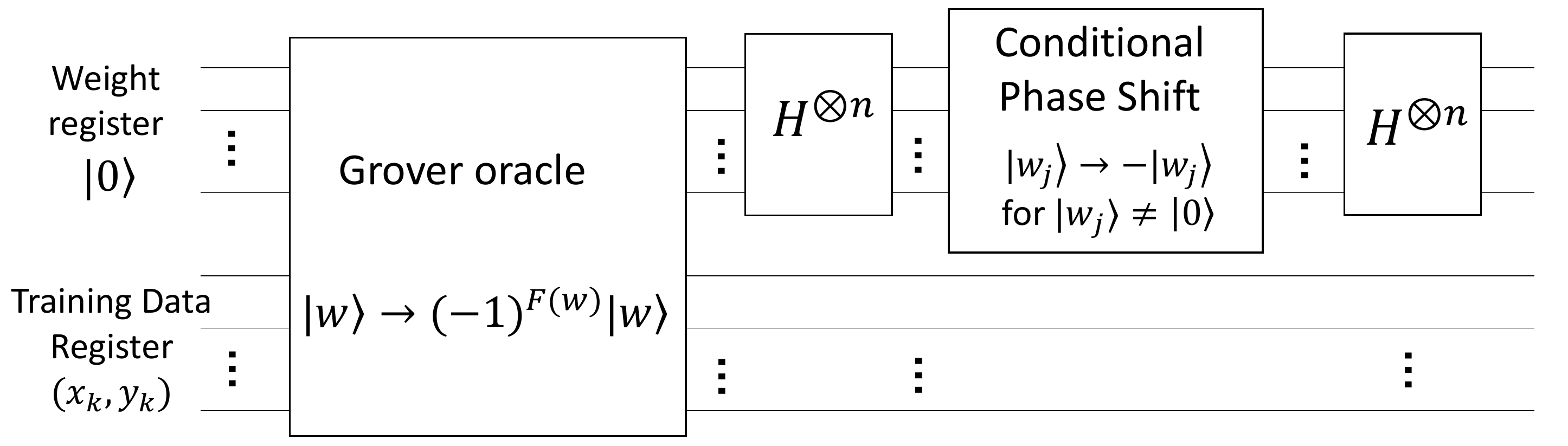}
	\caption{The schematic circuit for a single Grover iteration.}
	\label{fig:grover}
\end{figure}

\section{The quantum oracle circuit}\label{sec:oracle}

\begin{figure}
	\center
	\includegraphics[width=.8\columnwidth]{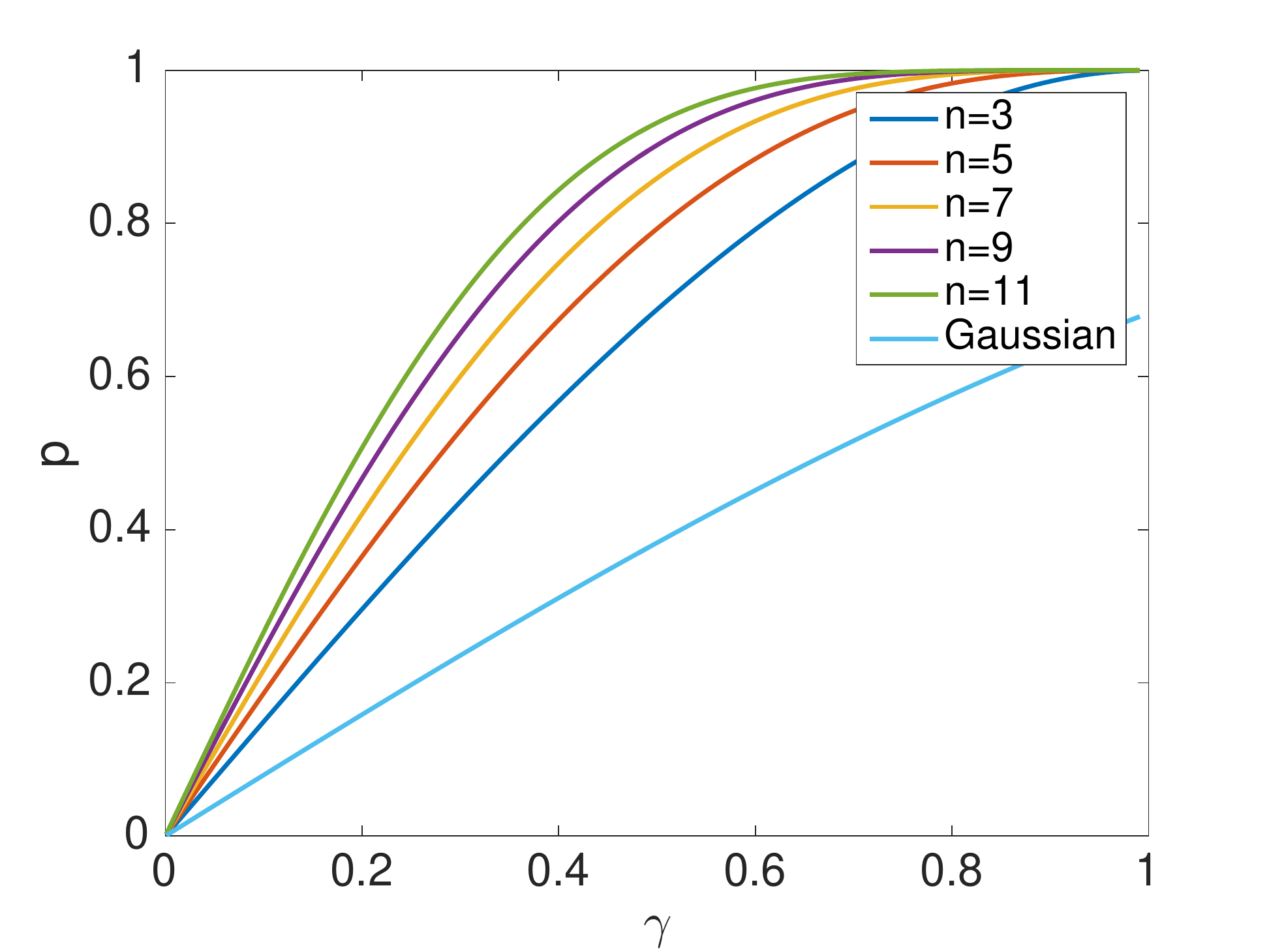}
	\caption{The comparison of success probability between Gaussian and uniform sampling schemes, where $n$ is the dimension of the feature space.}
	\label{fig:compare}
\end{figure}

In this section, we present our quantum circuit design for implementing the Grover oracle. The key part of the Grover-oracle circuit is the implementation of the Boolean function $F({\vec w})$, which involves the product and summation of digital numbers with signs.
We will implement them based on the Quantum-Fourier-Transform based circuits proposed in \cite{ruiz2014quantum} and analyze the required quantum resources. As shown in Fig.~\ref{fig:compare}, the circuit will be decomposed into units of quantum multiplier, quantum adder and quantum complementary (uncomplementary) circuits, which will be discussed in the following.

\begin{figure}
	\center
	\includegraphics[width=.8\columnwidth]{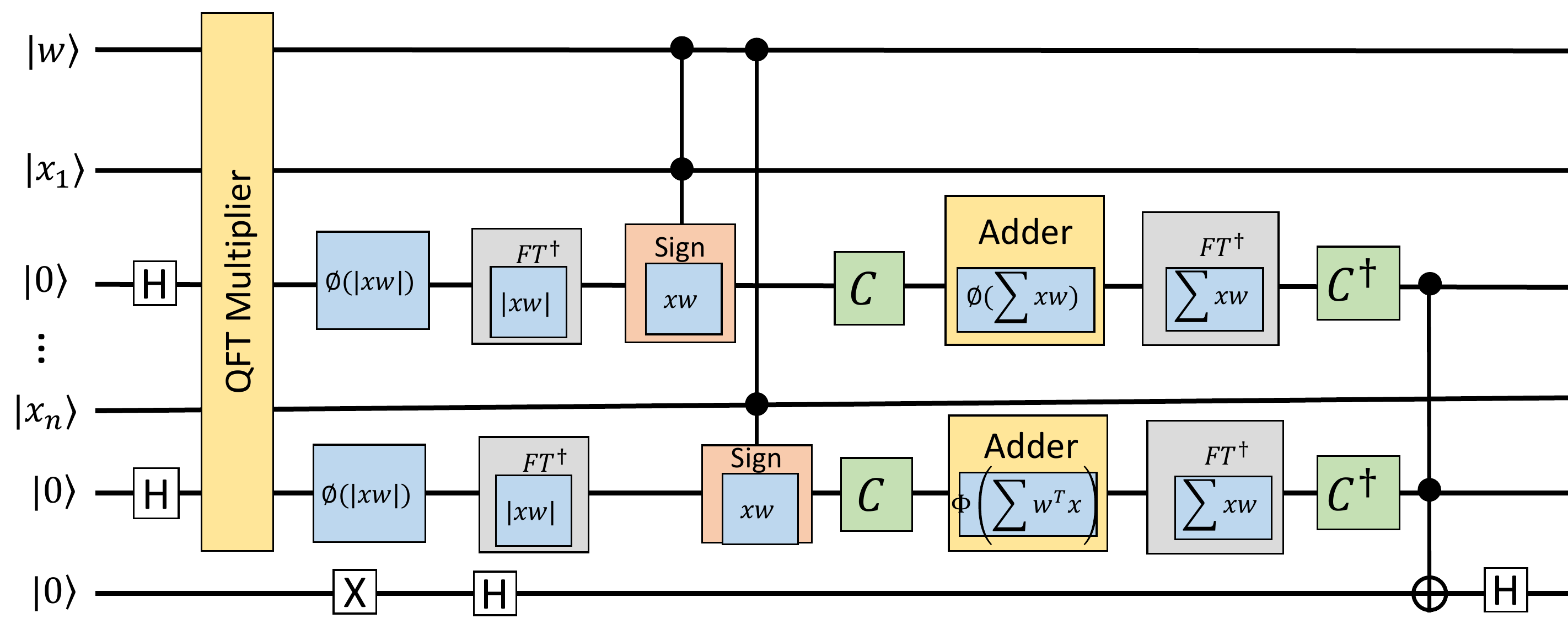}
	\caption{The quantum circuit for the Boolean function $f_{\vec w}(\vec x y)$. In this figure, the label $y_i$ has been merged into $x_i$.}
	\label{fig:arithmetics}
\end{figure}

\subsection{Quantum Multiplier}
The quantum multiplier computes the product of two scale numbers. This is required in the computation of the inner product $\vec w^T\vec x$, which is the sum of their entry-by-entry partial products. Suppose that the $k$-th entries of $\vec w$ and $\vec x$ are $a=(a_0a_1)_2$ and $b=(b_0b_1)_2$, respectively. First, we implement the QFT multiplier to compute $\phi(|ab|)$, the quantum Fourier transform of the absolute value $|ab|$ of $c=ab$. The sign of $ab$ will be handled later.

\begin{figure}[htbp]
	\center
	\includegraphics[width=0.8\linewidth]{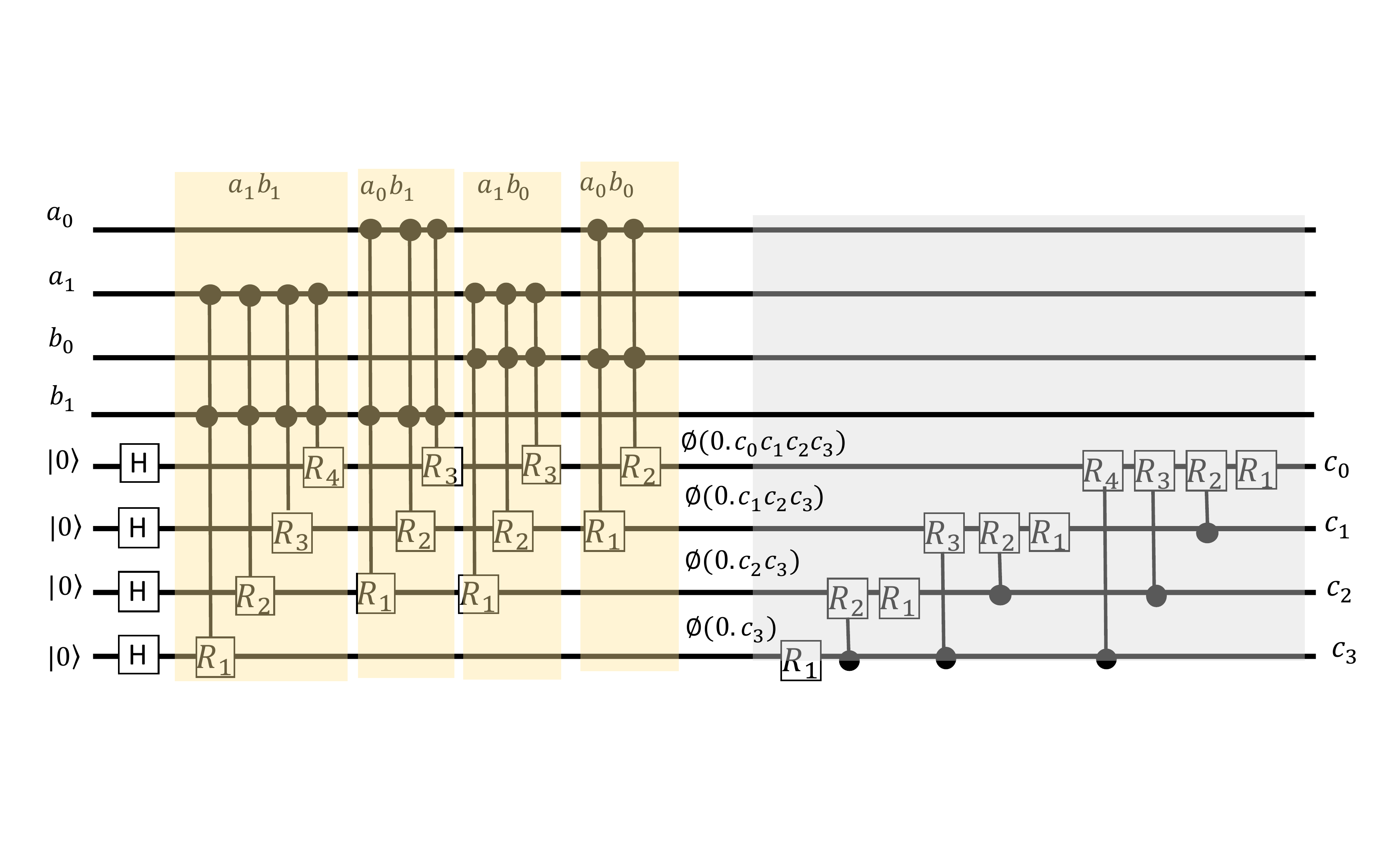}
	\caption{The quantum circuit for a QFT multiplier that calculates $c=a\times b$.}\label{fig:mul}
\end{figure}

If we use three digits to encode $c$, i.e., $c=c_0c_1c_2c_3$. Figure~\ref{fig:mul} shows how the multiplication is executed. The first block of the circuit computes the partial product by controlled rotation gates, where
$$R_k=\left(
        \begin{array}{cc}
          1 &  \\
           & e^{2\pi i/k} \\
        \end{array}
      \right),$$
and these gates produce the Fourier transformation of $c_0c_1c_2c_3$, $c_1c_2c_3$, $c_2c_3$ and $c_3$. After this, the following block performs an inverse Quantum Fourier transformation to obtain the result $c$.

The above circuit does not consider the sign of numbers, which is encoded by an additional single qubit $s$, i.e. $s=1$ for negative and $s=0$ for positive. If $a$ and $b$ are both signed numbers, the sign of their product $c=ab$ can be calculated as $s_c=s_a\oplus s_b$.

\subsection{Quantum Adder}

In the second step, we use the {QFT adder} to calculate the summation of all ${\vec w}_jx_j$, from which the classification result of a sample weight vector ${\vec w}$ can be obtained.

\begin{figure}
	\center
			\includegraphics[width=0.8\textwidth]{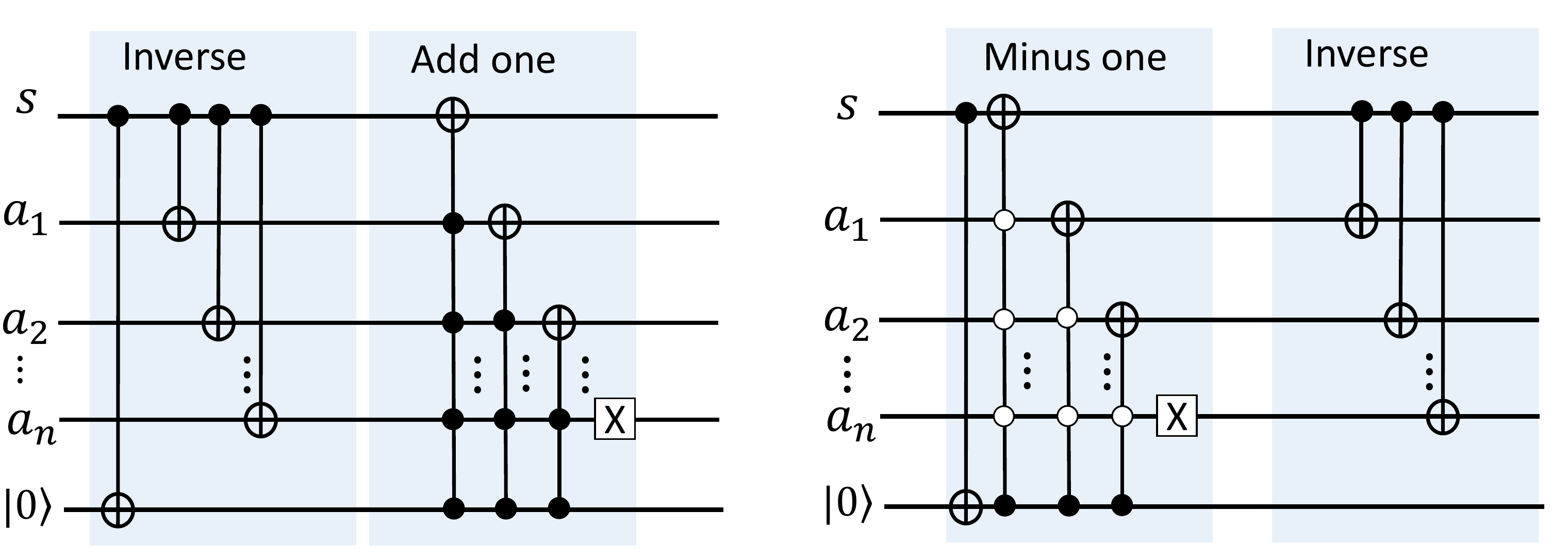}
	\caption{The quantum circuits for the complementary (left) and uncomplementary (right) of $n$-digit numbers.  }
	\label{fig:com}
\end{figure}

Because the adder can only deal with positive numbers, we use the quantum complement circuits to transform a negative number ${\vec x}$ into its complement code $[{\vec x}]_c$, an equivalent positive number (modular $2^n$).
As shown in Fig.~\ref{fig:com}a, we flip all qubits $a_1,\cdots,a_n$ when the sign $s=1$, otherwise nothing will be done. After adding the complementaries $[{\vec x}]_c+[y]_c=[{\vec x}+y]_c$, where $[{\vec x}]_c$ is the complementary code of ${\vec x}$, the result is then transformed back to a signed number through an uncomplementary circuit.

The summation of the products ${\vec w}_1{\vec x}_1,{\vec w}_2{\vec x}_2, ..., {\vec w}_nx_n$ is realize by a QFT adder \cite{ruiz2014quantum}. The adder uses a quantum Fourier transform to encode the digits of a number into the phase basis. Then, a phase gate is applied to transfer the digits to a phase shift that is equivalent to a modulo $d$ addition in that basis. In this way, the summation of two numbers is transform to the product of two phase gates. 

The final output of the Grover oracle corresponding to a given $|{\vec w}\rangle$ is obtained by the CNOT gates can sign the right ${\vec w}$ if $\bar{y}_i=y_i$ for all $i$.

\subsection{Resource assessment}
In our designed circuits, there are $N$ samples in an $n$-dimensional feature space, and each coordinate in this space needs $t\geq \log(\epsilon^{-1})$ qubits, where $\epsilon$ is the desired precision. So we need $Nnt$ qubits to encode these samples, which is the same as the number of classical bits used in a classical computer. Moreover, to take the advantage of QFT adder, we need additional $NnO(t)$ qubits to implement the inner products of ${\vec w}_ix_i$. As a result, the total number of qubits required by our designed quantum perceptron circuit is $(2Nnt+Nn)O(t)$.

As for the number of required universal gates, we can implement a QFT adder by $t^2$ gates, a QFT multiplier by $O(t^3)$ gates \cite{ruiz2014quantum} and a complement code circuit by $O(t)$ gates. The grover oracle needs to perform $N$ multiplications, the circuit for multiplier part consists of $O(N(tn)^3)$.
Therefore, we can implement our perceptron model with $O(Nt^3n^3)$ gates for $n$ features encoded in $t$ qubits as well as ${\vec w}$.

For classical algorithm, given only the ability to sample uniformly from the training vectors, the number of queries to $f_{\vec w}$ needed to find a training vector that the current perceptron model fails to classify correctly with probability $1-\epsilon\gamma^2$ is at most $O\left[\frac{N}{\gamma^2}\log\left(\frac{1}{\epsilon\gamma^2}\right)\right]$, where $\gamma$ means the training sequence is separable with margin $\gamma$. The computing complexity of this model is $O\left[\frac{N}{\sqrt{\gamma}}\log^{\frac{3}{2}}\left(\frac{1}{\epsilon}\right)\right]$\cite{wiebe2016quantum}.



\section{Experimental test wit IBM Q Cloud Quantum Computer}\label{sec:experiment}
\begin{figure}
	\center
	\includegraphics[width=.6\columnwidth]{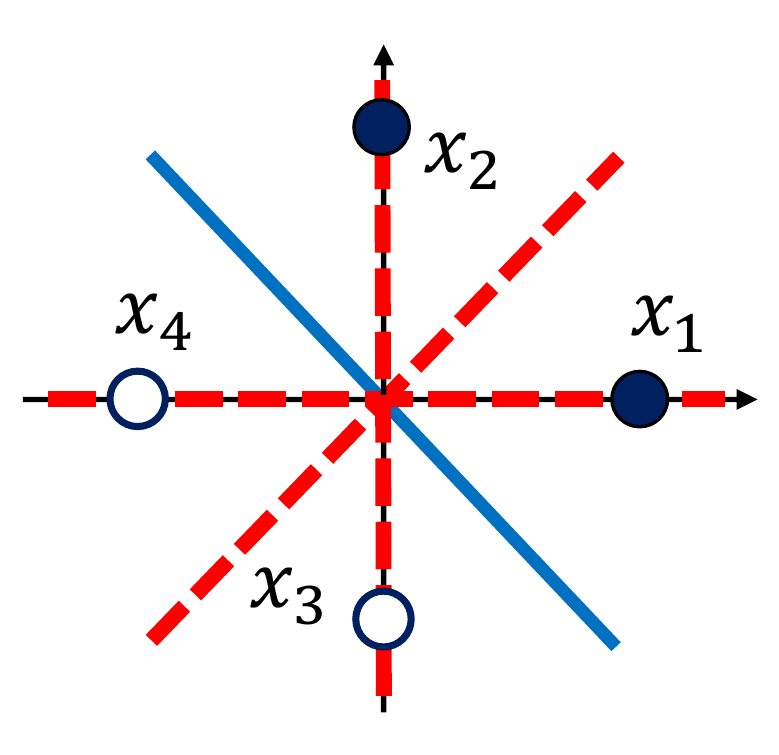}
	\caption{An example of a binary classifier with four samples in a two-dimensional feature space. Among the possible classifiers (the solid and dashed lines), only the solid line can correctly classify the data samples.}
	\label{fig:pro}
\end{figure}

To test the feasibility of the designed circuit, we propose an illustrative example shown in Fig.~\ref{fig:pro}, which contains four training data vectors in a two-dimensional space. We use 4 qubits to represent a signed number ${\vec w}$, each vector entry encoded by two qubits. With the Hadamard gate operation, we get a uniform superposition of 16 candidate weight vectors. However, these weight vectors only represent 9 candidates because zero and minus zero are the same.

\begin{figure*}[htbp]
	\centering
	\includegraphics[width=1\linewidth]{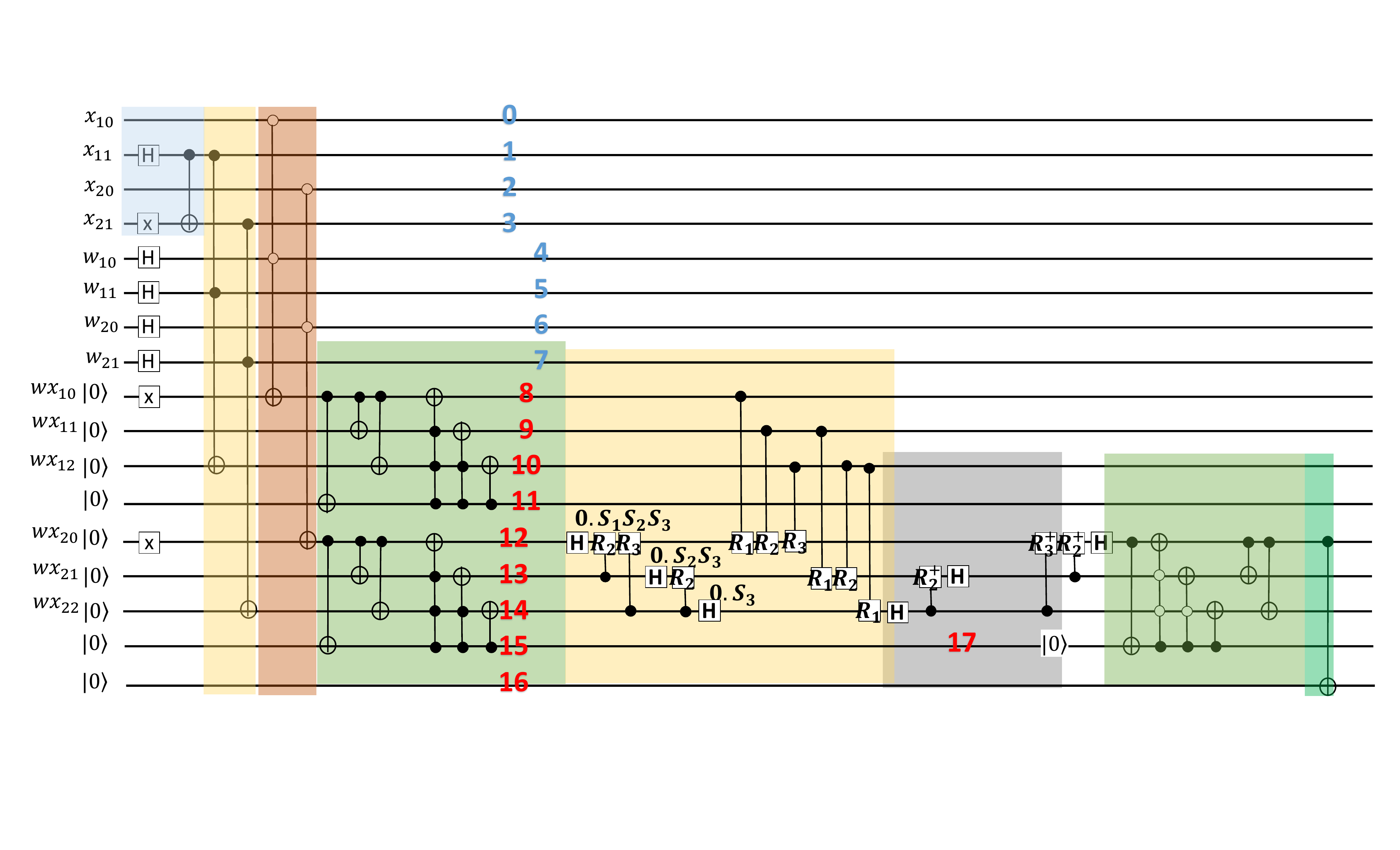}
	\caption{The entire circuit for the Grover oracle implementation of the example. In this figure, the label $y_i$ has been merged into $x_i$.}\label{fig:pg}
\end{figure*}

Among the 16 candidate vectors, it is easy to see that only one of them can correctly separate all four data samples, as shown in the solid line in Fig.~\ref{fig:pro}. According to the above analysis, we need at most repeat the Grover iteration for [$\frac{\pi}{4}\sqrt{\frac{{16}}{1}}$]$\simeq 3$ times. The entire quantum circuit for the Grover oracle is given by Fig.~\ref{fig:pg}, which uses 16 qubits to register the data and weight vectors, and intermediate results.

\begin{figure*}[htbp]
	\centering
	\includegraphics[width=1 \linewidth]{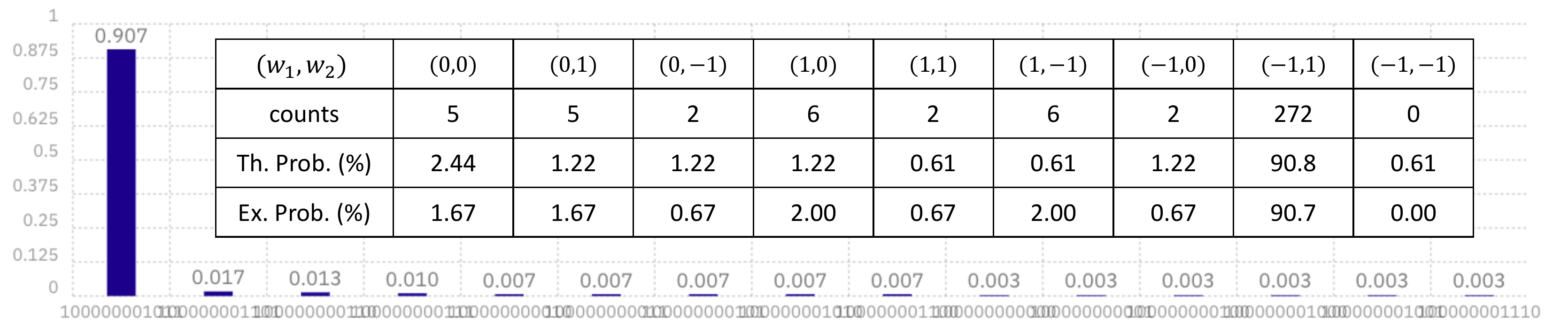}
	\caption{Experimental results of the quantum perceptron algorithm after three Grover iterations.}\label{fig:exp}
\end{figure*}

We applied the circuit to the IBM Q quantum computer \cite{ibmq}, and the result is shown in Fig.~\ref{fig:exp}. The experiments are repeated for 300 times, and in each experiment we do three Grover oracle iterations followed by he measurement of the weight vector. The statistics show that the amplitude of the correct classifier $(w_1,w_2)=(-1,1)$ is amplified to 90.7\%, very close to the theoretical prediction 90.8\%. This demonstrates the effectiveness of our designed circuit.

\section{Discussions}\label{sec:discussion}
In this paper, we present a detailed quantum circuit design for quantum perceptron model based on quantum Fourier transform based arithmetics. Analysis and experiments show that the circuit can successfully find the desired classifier with searching complexity reduced from $O(\frac{1}{\gamma^2})$ to $O(\frac{1}{\sqrt{\gamma}})$. However, note that this is at the price of increasing the number of register qubits, because each training data vector is encoded by an individual quantum register. 

In principle, it is possible, as proposed in the original paper \cite{wiebe2016quantum}, the training data can be loaded in parallel by a quantum memory in a superposition fashion, which will further reduce the circuit complexity to $O\left[\frac{N}{\sqrt{\gamma}}\log^{\frac{3}{2}}\left(\frac{1}{\epsilon}\right)\right]$ in limited qubits resources. However, we have not found a feasible circuit to implement this idea because it is hard to correctly assign phases to superposed weight register due to the complicated entanglement between the address register and the data register. Also, it is also unclear how to implement the quantum memory. Nevertheless, this approach is more intriguing for practical applications and should be explored in the future.

\section*{APPENDIX: Success probability of Grove search under uniform sampling}
Given $K$ samples hyperplanes ${\vec w}_1,\cdots,{\vec w}_K$ from a uniform distribution $U(0,1)$ has higher possibility than spherical Gaussian distribution $N(0,1)$.If the training data is composed of unit vectors, given that the margin of the training set if $\gamma$ there exist a hyperplane $u$ such that $y_iu^Tx_i>\gamma$ for all $i$. If ${\vec w}$ be a sample from $U(0,1)$ then lets compute what is the probability that perturbing the maximum margin classifier $u$ by amount ${\vec w}$ would still lead to a perfect separation. If we consider a data point ${\vec x}^*$ that lies on the margin, i.e. $y_iu^Tx_i=\gamma$, we are interested in the probability that $y_i({\vec w}+u)^Tx_i>0$ and $y_i({\vec w}+u)^Tx_i<2\gamma$. This is same as asking what is the probability that: $-\gamma<y_iw^Tx_i<\gamma$. Let us define $Z_i=y_iw^Tx_i$, as $||{\vec x}||=1$, we can regard ${\vec w}^Tx_i$ as the distance pbetween ${\vec w}$ and hyperplane ${\vec w}^Tx_i=0$. Since ${\vec w} \sim U(0,1)$, we can write the probability that $-\gamma<Z_i<\gamma$ as: For a n dimensional ball which $R=1$,
\begin{equation}
\begin{split}
p(-\gamma<Z_i<\gamma)=\frac{\int_{0}^{\gamma}r^{n-1}dz}{\int_{0}^{1}r^{n-1}dz}\\
\end{split}
\end{equation}
where $r=(1-z^2)^{n/2}$ and $n$ is the dimensional of features.
For odd $n=2k+1$,
\begin{eqnarray*}
&&p(-\gamma<Z_i<\gamma)\\
&&=C_k\sum_{m=0}^{k}\frac{(-1)^{m}k!}{m!(k-m)!(2m+1)}\gamma^{2m+1},
\end{eqnarray*}	
where $C_k={\sum_{m=0}^{k}\frac{(-1)^{m}}{2m+1}\frac{k!}{m!(k-m)!}}$

\bibliography{apssamp}

\end{document}